\documentclass[%
 reprint,
 amsmath,amssymb,
prb,
]{revtex4-2}
\usepackage[title]{appendix}
\usepackage{graphicx}
\usepackage{dcolumn}
\usepackage{bm}
\usepackage{gensymb}
\usepackage{float}
\usepackage{multirow}
\usepackage{tabularx}
\usepackage{amsmath}
\usepackage{color,soul}

\usepackage{physics}
\usepackage[colorlinks,allcolors=black,pdftex]{hyperref} 
\usepackage[capitalise]{cleveref}
\usepackage[dvipsnames]{xcolor}

\renewcommand{\b}[1]{{\boldsymbol{#1}}}

\begin{document}

\preprint{APS/123-QED}


\title{Structured light and induced vorticity in superconductors I: Linearly polarized light}

\author{Tien-Tien Yeh$^{1}$, Hennadii Yerzhakov$^{1}$,
Logan Bishop-Van Horn$^{2}$, 
Srinivas Raghu$^{2}$, 
Alexander Balatsky$^{1,3*}$}
\affiliation{$^{1}$Nordita, Stockholm University, and KTH Royal Institute of Technology, Hannes Alfvéns väg 12, SE-106 91 Stockholm, Sweden}
\affiliation{$^{2}$Department of Physics, Stanford University, Stanford, CA, USA}
\affiliation{$^{3}$Department of Physics, University of Connecticut, Storrs, Connecticut 06269, USA}

\begin{abstract}
We propose an approach to use  linearly polarized light  to imprint superconducting (SC) vortices. Within the framework of the generalized time-dependent Ginzburg-Landau equations we demonstrate the induction of the coherent vortex pairs that are  moving in phase   with electormagnetic wave oscillations. The overall vorticity of the superconductor remain zero throughout the cycle.   Our results  uncover rich  multiscale dynamics of SC vorticity and suggest new optical applications for various types of structured light. In departure from classical laser printing, the laser printing proposed here can be viewed as {\em quantum print} where we induce quantum excitations in the SC liquid. 
\end{abstract}

\maketitle

\section{\label{sec:intro}Introduction}

Since the invention of laser printing several decades ago, the challenge of printing in exotic materials has continuously evolved. Innovations include laser printing on bubble~\cite{zhao2024laser}, plasmonic nanoparticles~\cite{zhu2016plasmonic}, dielectric metasurfaces~\cite{zhu2017resonant}, nanostructure coloration~\cite{zhang2020bioinspired}, etc. 
With the development of structured light endowed with quantum numbers (QNs), the approach now has advanced to printing QNs onto the coherent quantum matter. Recent progress in advancing optical control  indeed demonstrates that idea or light transfer of QN is feasible. Examples include light spin angular momentum transfer to phonons~\cite{Basini2024Terahertz, davies2024phononic, luo2023large}, orbital angular momentum transfer to magnets~\cite{fujita2017encoding}, and spiral Higgs waves~\cite{mizushima2023imprinting}, Kapitza engineering in superconducting (SC) device controlling~\cite{yerzhakov2024laguerre}, induction of skyrmions~\cite{parmee2022optical} and polariton vortices~\cite{gnusov2023quantum, dominici2023coupled}. 
In superconductors, transferring various QNs from the light phase to the coherent quantum electron liquid holds the potential to open new pathways to the manipulation and control of superconducting states. In the standard laser printing one produces classical  printed objects -letters, images, by means of transfer of energy to paper.  We consider the set up where the magnetic field of the light can steer and induce quantized  vortices in electron fluid- SC. We view this process as a quantum version of printing - {\em quantum print}. The setup we use focuses on the light with frequency on the order of 0.1-10 THz (here we do include microwave in the possible source of light to be applied as well, hence the light term is used in general sense). 
The typical length scale of the light beam would be on the order of micron or larger. 
Hence we can look at the effects we describe as questions of stirring electronic hydrodynamics with light where we can use effective long wavelength description of the quantum fluid, i.e., hydrodynamic description applies. 

Specifically, in this work we simulate the behavior of superconducting film subject to incident light using time-dependent Ginzburg–Landau (TDGL) model with dynamical vector potential created by Gaussian and Laguerre-Gaussian beams~\cite{shen2019optical} to test this hypothesis. 

This work is divided into two parts. To set the stage, the first part, presented here, focuses on linearly polarized Gaussian beam. In the second part, we will elaborate on various kinds of QN printing to SC hydrodynamics, including spin angular momentum and orbital angular momentum, which will be published in the article ``Structured light and induced vorticity in superconductors II''~\cite{LG_TDGL_II}.

\begin{figure*}[!htbp]
    \centering
    \includegraphics[width=0.95\textwidth]{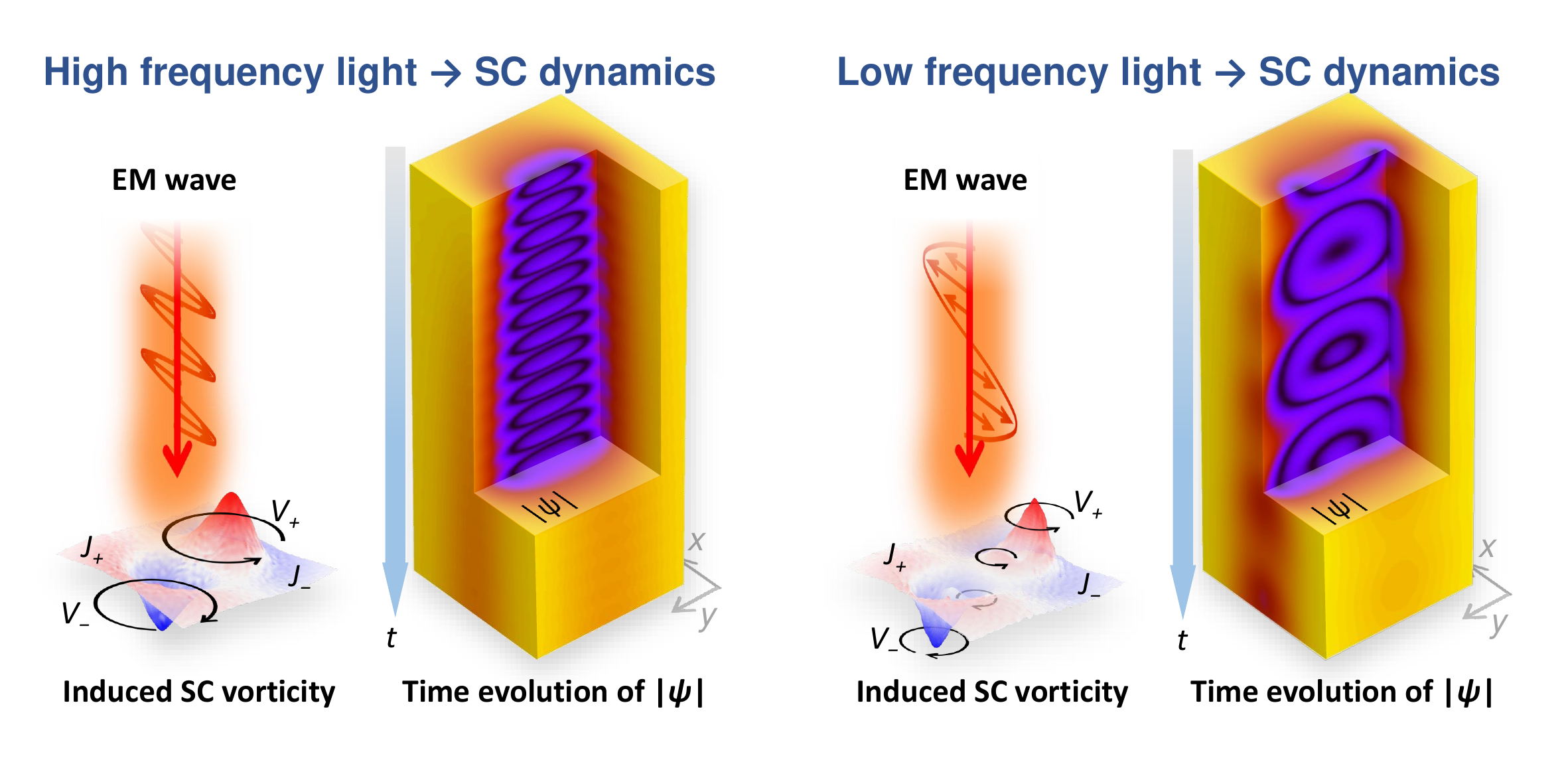}
    \caption{Linearly polarized light-induced SC vorticity and time evolution of order parameter $|\psi|$. The snapshots of vorticity show the light induced vortex-antivortex pair (labeled $V_+$ and $V_-$), supercurrent (labeled $J_+$ with positive vorticity and $J_-$ with negative vorticity).}
    \label{fig:abstract}
\end{figure*}

In this study, we report the predictions for the real-time motion coherent SC states with zero {\em total} QN transfer. The emphasis given since we will demonstrate that even though the net QN transfer is zero, there are regions in the beam spot where vortices will be generated 
due to time-dependent out-of-plane magnetic flux. In our simulations  of supercurrent vorticity, we track the dynamical interaction between light, superflow, and vortices, as illustrated in the schematic diagram of Fig.~\ref{fig:abstract}. In order to analyze the complicated  dynamics of superconducting state, we also examine the interplay between the length and time scales of the beam and superconducting fluid. We describe the effect of the  change  of coherence length and of the light spot size on the induced vorticity. We also demonstrate various dynamical regimes that emerge as the result of fast drive where damping is not efficient and we see evidence for light induced  oscillations (non-equilibrium regime)  to equilibrium relaxation where damping is efficient and the relaxation dominates.

Generation of vortices in superconducting state via transfer of the angular momentum of an electromagnetic beam has been proposed in Refs. \cite{yokoyama2020creation, plastovets2022all}. The  generation method focused on the superconducting condensate process during the quench of hot spot in this earlier work. We find that indeed structured light can induce the vortices even without relying on quench of thermally excited vortices. We also find  out that vortices can be generated without angular momentum transfer given the structured intensity of light provides locally non-zero magnetic flux through the plane normal to the propagation of the beam.

The paper is organized as follows. We describe the simulation method in~\cref{sec:method-1}, the light source and the simulation conditions in~\cref{sec:method-2}. Next, we uncover the typical imprinted order parameter pattern by linearly polarized light in~\cref{sec:fund1_1}, and show the evolution of linearly polarized light-induced supercurrent and vortices in~\cref{sec:osci}. In this section, we demonstrate long-lived cyclic vortex pair (VP) generation-recombination processes and non-repeatable VP generation-recombination process dependent on the optical frequencies and amplitude of optical induced flux.
Thereafter, we discuss the scale of printing of light on SC in 
~\cref{sec:fund1_2}. 
Finally, the discussion of the simulation results with the validity of TDGL model versus conventional SC materials, and proposing the experimental configuration with advanced light sources is presented in ~\cref{sec:discussion}.


\section{\label{sec:method}Method: Time-dependent Ginzburg–Landau theory} 

\subsection{\label{sec:method-1}Generalized time-dependent Ginzburg–Landau theory}

Our simulations are based on the generalized time-dependent Ginzburg–Landau theory (gTDGL). We adopt the dimensionless unit form of gTDGL, expressed as~\cite{kramer1978theory, watts1981nonequilibrium, bishop2023pytdgl, berdiyorov2014dynamics}
\begin{multline} 
\label{eq:TDGL}
\frac{u}{\sqrt{1+\gamma^2\left| \psi \right|^2}}
\left(\frac{\partial }{\partial t}+i\mu+\frac{\gamma^2}{2}\frac{\partial \left| \psi \right|^2}{\partial t}\right)\psi
\\ =(1-\left| \psi \right|^2)\psi+(\nabla -i\boldsymbol{A})^2\psi,
\end{multline}
where $\psi$ is the complex order parameter, consisting of the amplitude $|\psi|$ and phase $\theta_{s}$. The terms $(\nabla -i\boldsymbol{A})^2$ and $(\frac{\partial }{\partial t}+i\mu)$ represent the covariant Laplacian with vector potential $\boldsymbol{A}$ and the covariant time derivative with scalar potential $\mu$, respectively. 
The parameter $\gamma=2\tau_{E}\Delta_{0}$ represents the influence of inelastic scattering with scattering time $\tau_{E}$, and $\Delta_{0}$ is the superconducting gap in the absence of external perturbations at temperature $T$. The constant $u$ is the ratio of the relaxation times of the order parameter and current. For a SC with spherically symmetric Fermi surface in a normal state, $u=\pi^4/14\zeta(3)\approx5.79$~\cite{watts1981nonequilibrium}. 
The electric potential $\mu$ abides the Poisson equation 
\begin{equation}
\label{eq:mu}
\nabla^{2} \mu = \nabla \cdot \left( \boldsymbol{J_{s}} - \frac{\partial \boldsymbol{A}}{\partial t} \right) = \nabla \cdot \text{Im} [\psi^{*} (\nabla - i \boldsymbol{A}) \psi ] - \frac{\partial ( \nabla \cdot \boldsymbol{A})}{\partial t} ,
\end{equation}
and the total current density reads as
\begin{equation} 
\label{eq:J}
\boldsymbol{J}=\boldsymbol{J_{s}}+\boldsymbol{J_{n}}=\text{Im}[\psi^{*} (\nabla -i\boldsymbol{A})\psi] - \nabla\mu - \frac{\partial \boldsymbol{A}}{\partial t}.
\end{equation}

In~\cref{eq:TDGL,eq:mu} and throughout the paper, the temperature is implicitly included in the unit of time and length scale, which is encoded by the parameter $\epsilon(\boldsymbol{r})=1-T(\boldsymbol{r})/T_c(\boldsymbol{r})$ with the local transition temperature $T_c(\boldsymbol{r})$. We measure distance in units of the superconducting coherence length $\xi(T)=\xi_0/\sqrt{\epsilon}$, time in units of $\tau_{GL}(T) =  \mu_{0} \sigma \lambda_{L}(T)^2 = \mu_{0} \sigma \lambda_{L,0}^2 / \epsilon = \tau_{GL,0} / \epsilon$, where $\sigma$ is the conductivity in the normal state and $\lambda_{L}(T) = \lambda_{L,0} / \sqrt{\epsilon}$ is the London penetration depth, magnetic field in units of $B_{c2}(T) = \epsilon B_{c2,0} = \epsilon \Phi_{0}/2\pi\xi_0^2$, vector potential in units of $A_0(T)=\xi(T) B_{c2}(T)$, current density in units of $J_0(T)  = \xi(T) B_{c2}(T)/ \mu_0 \lambda_{L}(T)^2$, and scalar potential $\mu$ in unit of $V_0(T)=\xi(T) J_0(T)/\sigma(T)$. The units $\xi, \tau_{GL}, \lambda_{L}$, etc are considered at temperature $T$ throughout the paper. The restriction of $T$ is discussed in~\cref{sec:applicability}.

We shine light normally onto a square-shaped quasi-two-dimensional superconducting thin film with thickness $d$, assumed to be much less than $\xi$, and length $L$. In a quasi-2D case, the effective penetration depth is played by the Pearl length $\Lambda = \lambda_{L}^2/d$~\cite{pearl1964current}.

Throughout this work, we use $\gamma=10$. In the simulations, we neglect screening effects choosing the thickness to be sufficiently small to result in a large effective penetration depth $\Lambda$. The boundary conditions and other details are discussed in Refs.~\cite{kramer1978theory, watts1981nonequilibrium, bishop2023pytdgl}. ~\cref{fig:schm}(c) demonstrates the schematic diagram of gTDGL simulations.
The dynamical $\boldsymbol{A}$ and the aforementioned parameters of SC thin film are set as input for the solver in the Python-based open-source pyTDGL package~\cite{bishop2023pytdgl}, which carries out numerical solution of gTDGL equations. Finally, the simulation yields the order parameter and currents density as an output.


\subsection{Time-dependent vector potential}
\label{sec:method-2}

In this work, the light source is linearly polarized in $x$-direction Gaussianly structured beam incident along $z$-direction onto a SC thin film positioned at $z=0$. Thus, the vector potential is given by 
\begin{equation} 
\b{A_{EM}}(\boldsymbol{r}, t)=\b{E}(\boldsymbol{r}, t)/i\omega_{EM}
\label{eq:A_EM}
\end{equation}
 with
\begin{equation} 
\label{eq:E_EM}
\boldsymbol{E}(\boldsymbol{r}, t) = E_{0}(\boldsymbol{r}) e^{i \omega_{EM} t - i k z} \boldsymbol{\hat{e}_x},
\end{equation}
where $\omega_{EM}$ and $k$ are angular frequency (hereinafter referred to as frequency) and wavenumber of the light, respectively. The parameter $\b{r}$ is the radius vector in the plane of the film with the origin at the beam's center, and $E_{0}(\boldsymbol{r}) = E_{amp} e^{-\abs{\boldsymbol{r}}^2/w_{0}^2}$ is Gaussianly distributed amplitude, where $E_{amp}$ is the amplitude at the center of the beam, and $w_{0}$ is the radius of the spot of the beam. The period of EM wave $\tau_{EM}$ is equivalent to $2\pi/\omega_{EM}$. 

\begin{figure*}
    \centering
    \includegraphics[width=0.9\textwidth]{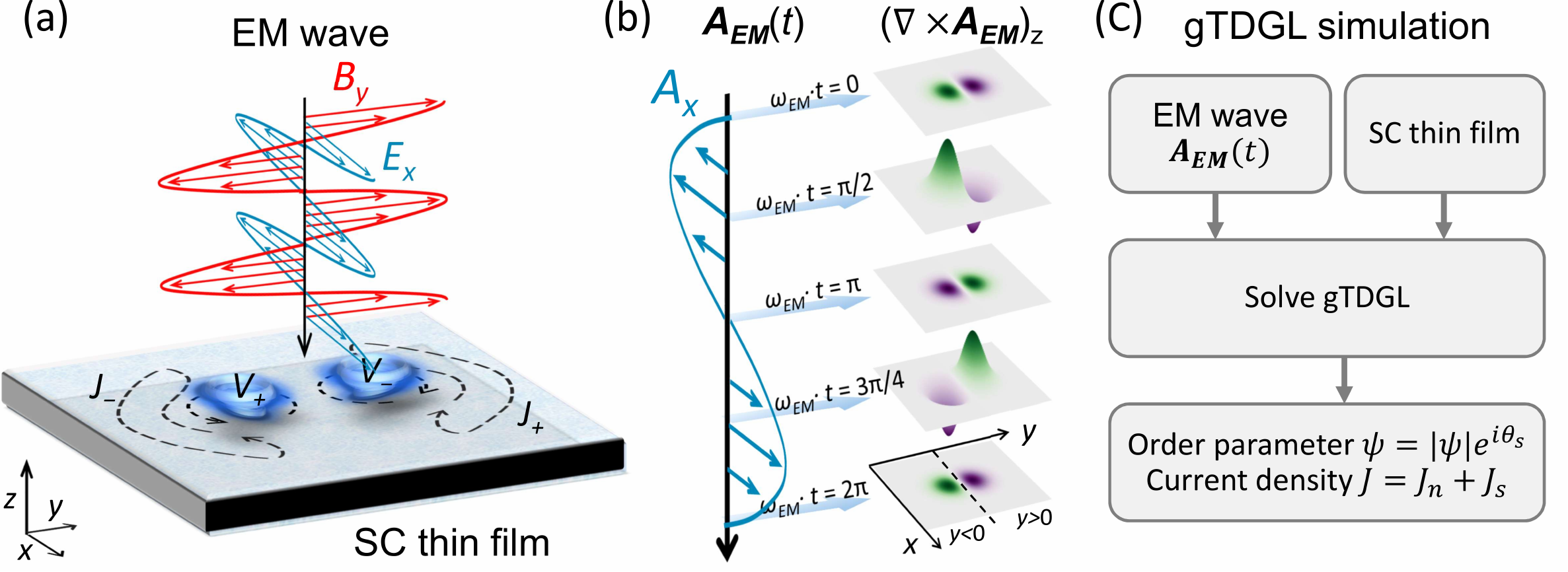}
    \caption{(a) Schematic diagram of applied EM wave and SC thin film. (b) $A_{EM}$ and profile of the corresponding $(\nabla\times\boldsymbol{A_{EM}})_z$ at different time frames. (c) Flowchart of the gTDGL simulation.}
    \label{fig:schm}
\end{figure*}

A simple but key observation is that gradient of the vector potential gives rise to a locally non-zero oscillating $z$-component of the magnetic field 
\begin{align}
\label{eq:B_z}
    B_z(\boldsymbol{r},t) = (\nabla\times\boldsymbol{A_{EM}})_z = -\frac{\partial A_{EM,x}}{\partial y} = \\ \nonumber
    2 \frac{E_{amp}}{i \omega_{EM}} \frac{y}{w_0^2} e^{-\abs{\boldsymbol{r}}^2/w_0^2} e^{i \omega_{EM} t}
\end{align}
with maximums of its absolute value located at $y_0 = \pm w_{0}/\sqrt{2}$ on the $y$-axis. Due to antisymmetry of $B_z$ in $y$, the total flux through the film is zero. However, locally flux is not zero (see~\cref{fig:schm} (b)), and the values of the flux through the half-planes $y \lessgtr 0$ are
\begin{align}
\label{eq:Phi_B}
    \Phi_{y \lessgtr 0} = \pm \Phi_B = \pm \frac{\sqrt{2} E_{amp}}{i \omega_{EM}} w_0 e^{i \omega_{EM} t}.
\end{align}
When the amplitude of $\Phi_B \gtrsim \Phi_0$, we might expect to see generation and annihilation processes of vortices due to the time-dependent magnetic flux.

As we use a number of symbols, for convenience of the reader, we compile Table~\ref{tab:symbol}, which provides a comprehensive list of these symbols.

\begin{table}[h!]
\caption{List of symbols.}
\begin{tabularx}{0.5\textwidth}{lX}
\hline
Quantity &  Explanation  \\ 

$|\psi|$ & Amplitude of complex SC order parameter $\psi$ \\
$\theta_s$ & Phase of complex SC order parameter $\psi$ \\
$\boldsymbol{A_{EM}}$ & Vector potential of EM wave \\
$\tau_{GL}$ & Characteristic time of TDGL equation at $T$\\
$\tau_{EM}$ & Period of EM wave \\
$\tau_{healing}$ & Relaxation time of healing process of $\psi$ \\
$\omega_{GL}$ & Characteristic angular frequency of TDGL at $T$ \\
$\omega_{EM}$ & Angular frequency of EM wave \\
$\xi$ & SC coherence length at $T$ \\
$\lambda_L$ & London penetration depth at $T$ \\
$\Lambda$ & Pearl length, also represent penetration depth for SC thin film at $T$\\
$L$ & Length of square size SC sheet in simulation \\
$\boldsymbol{r}$ & Vector between position and center of light spot \\
$w_0$ & radius of light with Gaussian distribution \\
$\lambda_{EM}$ & wavelength of EM wave \\
$J_0$ & Unit of current density in gTDGL\\
$\boldsymbol{J}$ & Total current density \\
$J_+$/$J_-$ & Current density with positive/negative vorticity  \\
$V_+$/$V_-$ & Vortex with positive/negative vorticity \\  
$\omega_{\nu,s}$ & Vorticity of SC current density \\
VP$_j, j=1,2,...$ & The $j$-th pair of vortex pair \\
$E_{amp}$ & Amplitude of electric field at center of Gaussian distribution and peak value of period of EM wave\\
$\boldsymbol{E}$ & The electrical field of Gaussian beam \\  
$B_z$ & The out-of-plane magnetic field \\
$B_{c2}$ & Critical magnetic field of Type II SC at $T$ \\
$\Phi_B$ & The out-of-plane magnetic flux \\ 
$\Phi_0$ & Magnetic flux quantum \\
$\sigma$ & Conductivity of normal state at $T$ \\
$r_n$ & Resistivity of normal state \\ 
$v_F$ & Fermi velocity \\ \hline
\end{tabularx}
\label{tab:symbol}
\end{table}

We emphasize that the simulations are carried out in thin films, i.e., $d \ll \xi$, which allows to neglect variations of the order parameter, current, and vector potential in $z$-direction. 
Additionally, as we neglect the screening effects, the Pearl length is supposed to be much larger than the sample size or the effective size of the light spot. 
Unless specified differently, in simulations we consider SC thin films with $L=20 \xi$, $d=0.02 \xi$ and $T=0.99T_c$, which for $\xi=100$ nm corresponds to $L=2$ $\mu$m and $d=2$ nm. Further details about $w_0$ and screening length of materials are discussed in~\cref{sec:optic}.


\section{Results}
\label{sec: Results}

Here, we begin by presenting the typical imprinted vorticity distribution and 
SC motion with continuous light source, and giving the picture how coherent state react dependent on the amplitude of light. Furthermore, we focus on investigating the imprinted profile of SC with different conditions of light source.


\subsection{\label{sec:fund1_1}Optical imprint and vorticity distribution} 

\begin{figure*}
    \centering
    \includegraphics[width=0.9\textwidth]{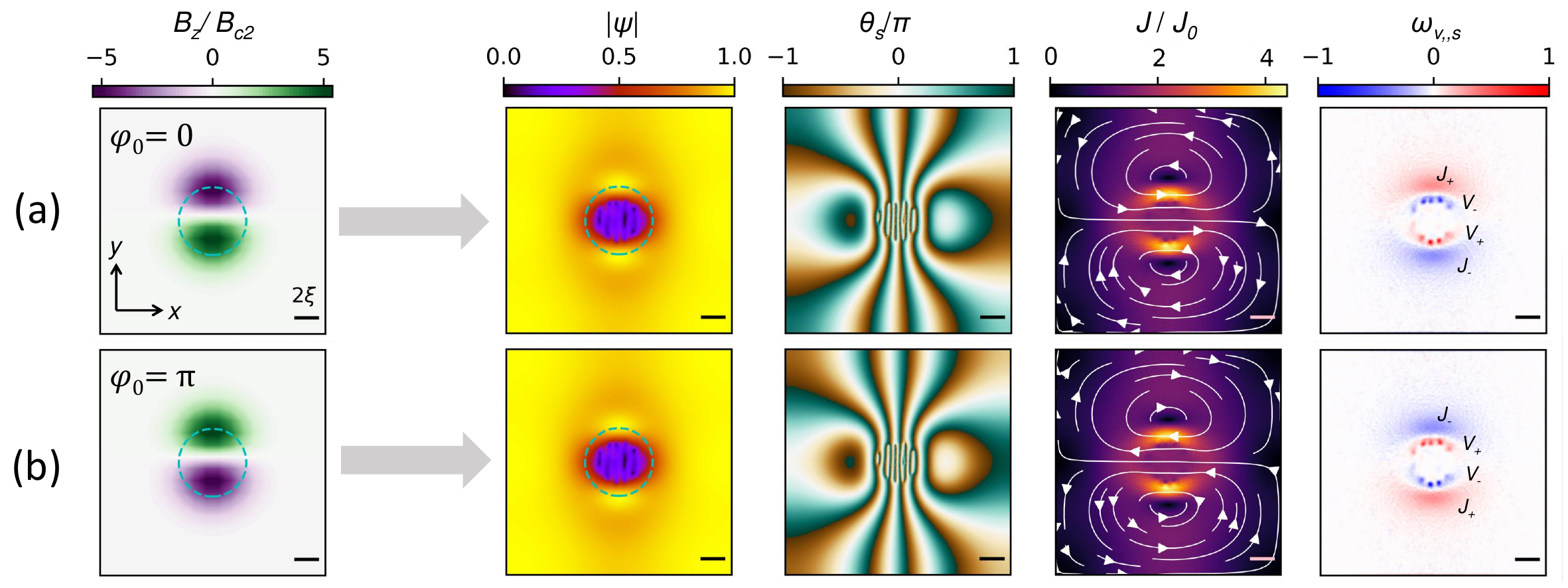}
    \caption{The SC state excited by the different sign of $B_z$. The sign of $B_z$ is tuned by add the phase $\phi_0$ into Eq.~\ref{eq:E_EM}. (a) ($\phi_0=0$, first row) and (b) 
    ($\phi_0=\pi$, second row) represent the magnetic flux from the same light source with 0 and $\pi$ phase shift, respectively. The profiles of $B_z/B_{c2}$ are shown on left side of gray arrows. The corresponding imprint distribution of the SC state show at right side of gray arrows, including $|\psi|$, $\theta_s$, $J/J_0$, and vorticity.
    The green dashed circles and inset axis at the left figures represent the beam spot and direction of coordinate. The notations $J_+$, $J_-$, $V_+$ and $V_-$ are consistent with those in Table~\ref{tab:symbol}. Both (a) and (b) show 5 VPs in the snapshots.
    }
    \label{fig:distrib}
\end{figure*}

We first demonstrate the imprinting of vortices by Gaussian linearly-polarized light in~\cref{fig:distrib}. It represents a snapshot of the simulations at time $2 \tau_{GL}$ for a beam of light at frequency $\omega_{GL}/8$ and $w_0 = 3 \xi$ which produces $\Phi_B$ of amplitude $16 \Phi_0$. 
\cref{fig:distrib} (a) anb (b) are produced for the initial phases of the EM wave $\phi_0=0$ and $\phi_0 = \pi$, respectively. 
We show in Fig.~\ref{fig:distrib} that the phase of the magnetic field is directly controlling the formation of vortices of certain orientation: ~\cref{fig:distrib}(a) $B_z$ and ~\cref{fig:distrib}(b) opposite sign of $B_z$ compared to ~\cref{fig:distrib}(a).

As expected, the density plots for the magnitude of the order parameters are the same, while the current densities and the winding numbers of vortices change sign. 
In the profile of vorticity, the distributions of $V_+$, $V_-$. $J_+$, $J_-$ are opposite between subplot (a) and (b). The appearance of vortices are induced by the excessive $\Phi_B(t)$.
In these simulations, the maximum induced current density exceeds the critical current density $J_c \approx 0.48 J_0$ (see~\cref{app:Jc}), which leads to significant suppression of the amplitude of the order parameter in the region of the beam's spot. This imprint pattern is presented by the distribution vorticity of vortex generation.

\subsection{\label{sec:osci}Dynamics of light-induced supercurrent and vortex}

\begin{figure*}
    \centering
    \includegraphics[width=0.9\textwidth]{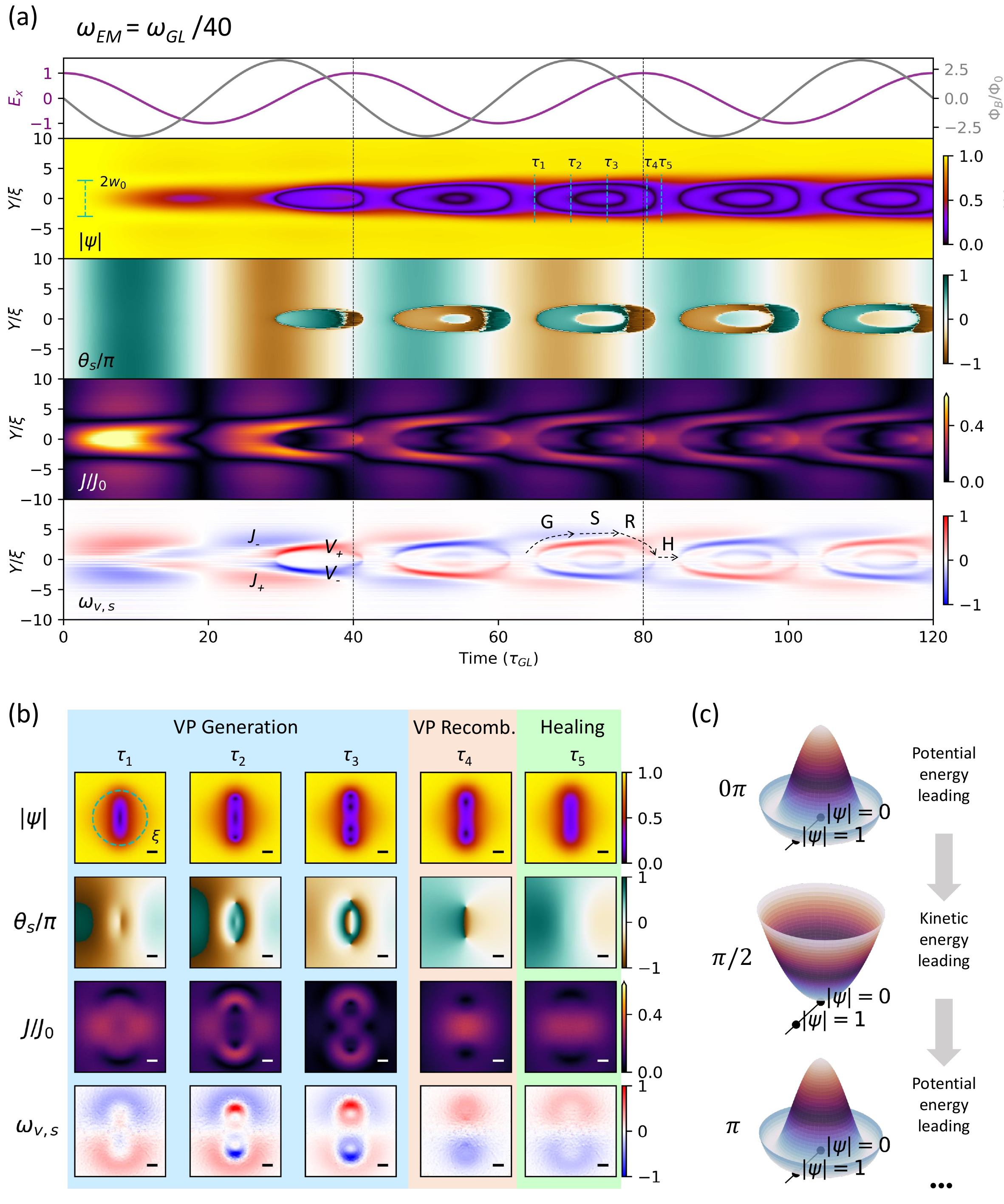}
    \caption{
    (a) The SC time evolution driven by light source with $\omega_{EM} = \omega_{GL}/40$. In the top row, the purple curve represents the normalized $E_0$ referenced the left axis, and gray curve is the out-of-plan flux $\Phi_B/\Phi_0$ referenced right axis. The second to fourth subplots show the vertical slice at middle of sample for $|\psi|$, $\theta_s/\pi$ and $\omega_{\nu,s}$. In the subplot of $|\psi|$, the left green dashed line marked $2w_0$ is the size of spot, and right dashed lines marked $\tau_1$ to $\tau_5$ represent selected moments corresponding to different stages of VP evolution, detailed in (b). In the $\omega_{\nu,s}$ subplot, the notations $J_+$, $J_-$, $V_+$ and $V_-$ are consistent with those in ~\cref{tab:symbol}. The dashed curves labeled G, S, R, H denotes the duration of the generation, quasi-steady state (dwell time of VP$_1$), recombination, healing processes, respectively. The black dashed lines cross all subplots at 40 and 80 $\tau_{GL}$ mark the interval of EM wave period.  
    The complete dynamical results of $|\psi|$, $\theta_s$, $J$, and $\omega_{\nu,s}$ also demonstrate in the supplementary materials~\cite{supp3}.  
    (b) Snapshot of $|\psi|$, $\theta_s$, $J$, and $\omega_{\nu,s}$ at different stages of the VP life time. 
    (c) Schematic diagram of the cycling of free energy density. 
    }
    \label{fig:osc_slow}
\end{figure*}

In~\cref{fig:osc_slow}, we illustrate the results of simulations for a Gaussian beam of spot size $2 w_0 = 6 \xi$ with $\omega_{EM}=\omega_{GL}/40$ and amplitude $E_{amp} = 2 A_0 \omega_{GL}$, which correspond to the amplitude of $\Phi_B = 3.2 \Phi_0$. 
The choice of $\omega_{EM}$ here is dictated by the effective relaxation time of the healing of $\abs{\psi}$ after recombination of vortices, which is about $31 \tau_{GL}$ (see ~\cref{app:relaxation_time}).
The center of the beam hits the center of the film, where we locate the origin of coordinates. After initial delay, we observe generation and recombination of vortex-antivortex pairs (VPs). In~\cref{fig:osc_slow} (b), We illustrate this process in the snapshots of the order parameter amplitude, phase, magnitude of the current density, and SC current vorticity $\omega_{\nu,s}$ at $t=\tau_1,...,\tau_5$ in the second cycle of the EM wave (see~\cref{fig:osc_slow} (a)) . As can be seen, the VP motion occurs along the $y-$direction. This allows us to illustrate the VP evolution in~\cref{fig:osc_slow} (a), where in the top row we plot time-dependence of the electric field and $\Phi_B$, in the second and third rows we plot the magnitude and the phase of the order parameter along the line of the VP motion ($x=0$ in this case) as a function of time, and in the fourth and the bottom row we plot the magnitude of the full current and the vorticity of the SC current, respectively, along the line $x=0$ as a function of time.

In the first half of cycle, $\abs{\psi}$ start decreasing first at $y=0$. Since $B_z$ is zero at $y=0$, we explain this by that that locally current density exceeds the critical current density, i.e., $J/J_c>1$. Subsequently, one VP (VP$_1$) is generated during the second half-cycle of the EM wave. In the third half-cycle, at some time after VP$_1$ is generated, $\abs{\psi}$ in the center of the spot decreases significantly, which can be interpreted as a precursor for the second VP (VP$_2$). VP$_2$ fully appears in the fourth EM wave half-cycle, after which this process becomes (almost) periodic. The windings of the phase around vortices, clearly visible in~\cref{fig:osc_slow} (b), are reflected as jumps in the phase in~\cref{fig:osc_slow} (a).

Let us discuss the evolution of VPs in more details on the instance of their motion that occur during the fourth half-cycle of the EM wave. The outer pair, VP$_1$, appears at the moment of time $\tau_1$, when $\Phi_B \approx 1.9 \Phi_0$, and the vortices in the pair moves relatively fast towards positions at $y_{\pm} \approx \pm 0.7 w_0$ (whereas maximums of $B_z$ are located at $y_0 = \pm w_0/\sqrt{2} \approx 0.71 w_0$) until time $\tau_2$ ($\tau_2 - \tau_1 = 5 \tau_{GL}$). After that vortices in this pair stay around positions at $y_{\pm}$ until time $\tau_3$. Shortly after time $\tau_2$, when $\Phi_B$ reaches $3.2 \Phi_0$ after passing its maximum, the second pair, VP$_2$, is generated and follows similar path as VP$_1$ until time $\tau_3$. After time $\tau_3$, vortices and antivortices in both pairs start moving towards each other marking the beginning of the recombination process. Outer VP$_1$ annihilates somewhat after the flux reaches zero value. We also note that the recombination process is associated with the change of sign of the supercurrent vorticity away from vorticies ($J_{\pm}$ in~\cref{fig:osc_slow} (a)). Overall, there is a significant reduction of order parameter around center of the beam. While this happens due to complicated dynamics of the current distribution, this partially can be explained noting that  $\b{A}^2 \psi$ term in~\cref{eq:TDGL} can be interpreted as the modification to the mass term in the Lagrangian from which gTDGL equaion is derived, effectively decreasing local $T_c$ (see ~\cref{fig:osc_slow} (c)).

\begin{figure*}
    \centering
    \includegraphics[width=0.9\textwidth]{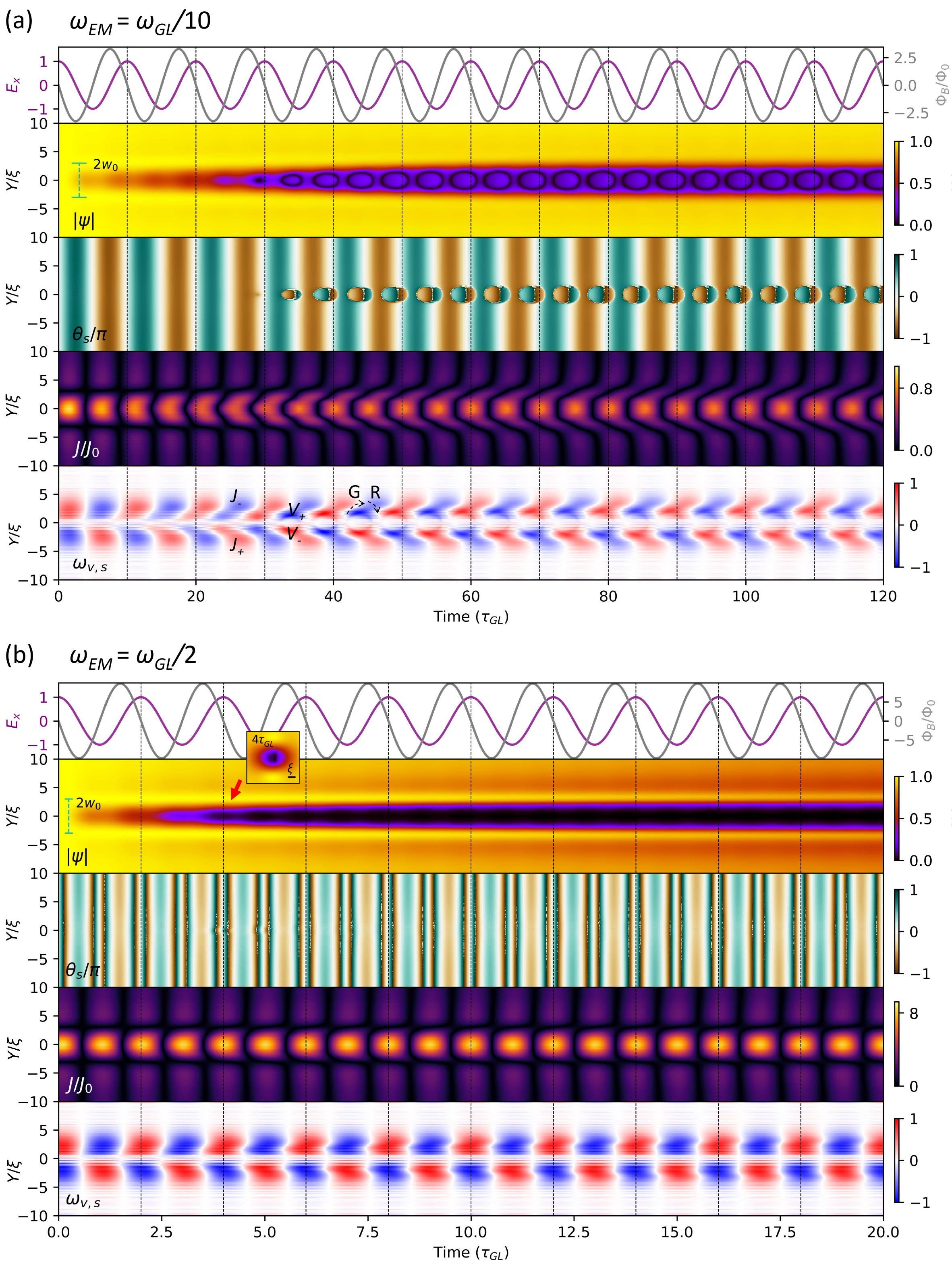}
    \caption{Similar simulations as~\cref{fig:osc_slow} but with highter $\omega_{EM}$: (a) $\omega_{EM}=\omega_{GL}/10$ and (b) $\omega_{EM}=\omega_{GL}/2$. The red arrow insets in (b) indicate a VP life cycle, and the inset of $\abs{\psi}$ shows the profile of VPs at $4 \tau_{GL}$. The simulation condition of $E_{amp}$ in (a) is 4 times larger than the case of Fig.~\ref{fig:osc_slow}, and in (b) is 60 times larger. The complete dynamical results of $|\psi|$, $\theta_s$, $J$, and $\omega_{\nu,s}$ also demonstrate in the supplementary materials~\cite{supp4} and ~\cite{supp5}. 
    }
    \label{fig:osc_fast}
\end{figure*}

In~\cref{fig:osc_fast} (a), we show the results for the simulations in which we changed the EM wave frequency to $\omega_{EM}=\omega_{GL}/10$. To keep the same amplitude of the flux, we correspondingly increased the amplitude of the EM wave by four times, i.e., $E_{amp}=8 A_0 \omega_{GL}$. This time we observe generation of only one VP, which first happens during the seventh half-cycle of the EM wave. Subsequently, it undergoes recombination, after which the process of the generation-recombination reoccurs almost periodically. A VP is generated almost at times when $\Phi_B$ reaches its maximum/minimum value of $\pm 3.2 \Phi_0$. After that the vortices in the pair move relatively fast for about $1.9 \tau_{GL}$ on distance about $0.45 w_0$ each and recombine significantly after the flux changes the sign when $\Phi_B = \mp 1.9 \Phi_0$. We note that during the generation-recombination cycle of the VP, the electric field growth from almost zero, reaches its maximum/minimum value, and goes to intermediate values without changing the sign at the recombination time. This is unlike the previous case, where at the generation time of VP$_1$ $E_x$ has intermediate values, then changes sign, and reaches values close to its maximum/minimum at the recombination time. While the data supports our conjecture that the generation of a VP is induced by sufficiently strong $z-$component of the magnetic field, this observation highlights the complex interplay of timescales set by the EM wave and gTDGL equation. 

When we further increase $\omega_{EM}$ to $\omega_{GL}/2$, there is no generation of VPs at all for the intensity of the beam which produces the same amplitude of $\Phi_B = 3.2 \Phi_0$. When the intensity increased such that amplitude of $\Phi_B$ equals to $9.6 \Phi_0$, we can see a cycle of VP generation-recombination process (see~\cref{fig:osc_fast} (b)), after which the light simply destroys SC locally leaving a ``scar'' in the SC order parameter.

\begin{figure*}
    \centering
    \includegraphics[width=0.9\textwidth]{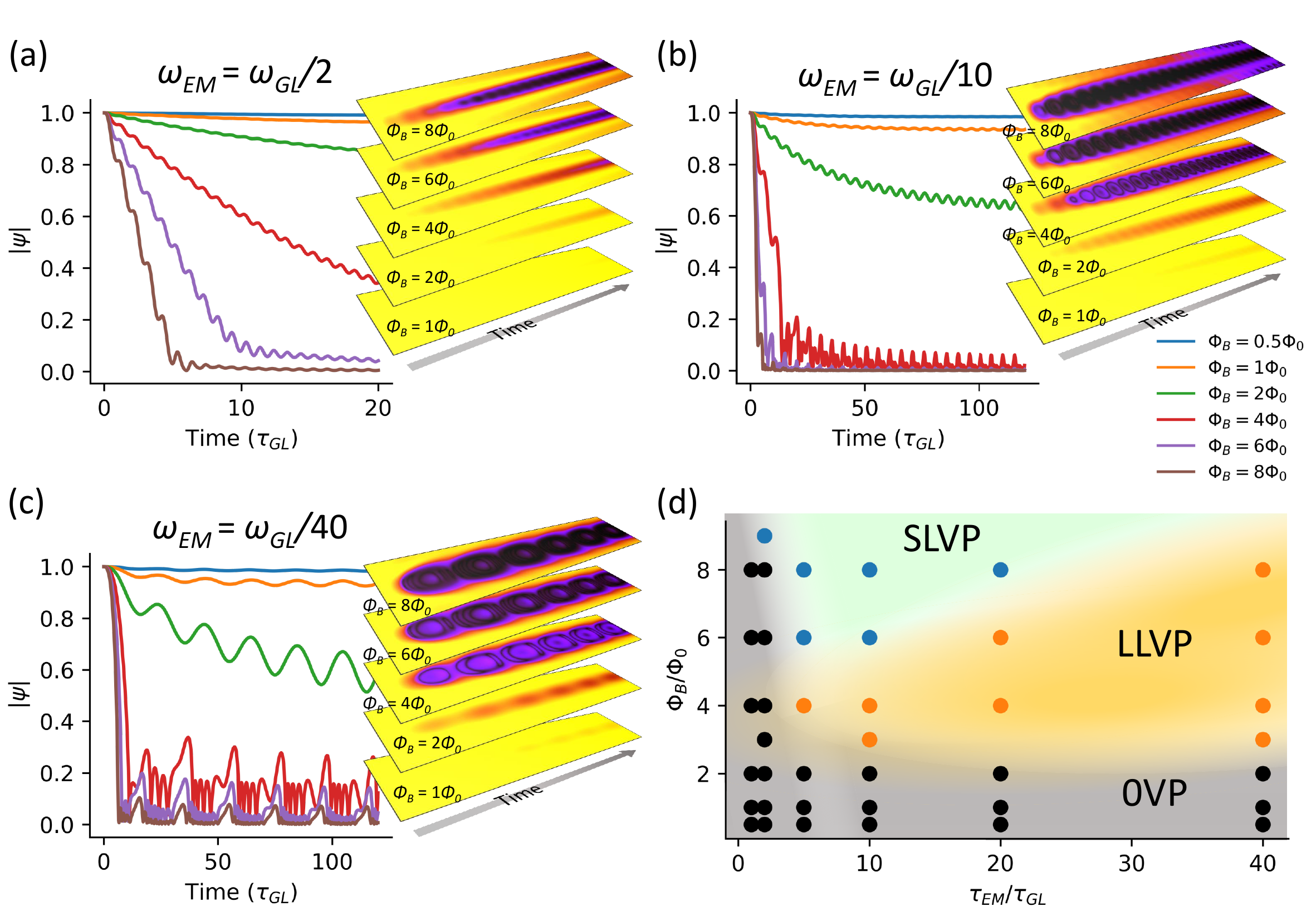}
    \caption{The time traces of $\abs{\Psi}$ at $x, y= (0,0)$ for (a) $\omega_{EM}=\omega_{GL}/2$, (b) $\omega_{EM}=\omega_{GL}/10$, and (c) $\omega_{EM}=\omega_{GL}/40$. The yellow inserts show the time-dependent profile along $x=0$ with different $\Phi_{B}$. (d) The results of $\tau_{EM}$ and $\Phi_{B}$ dependent on state of VP cycle. The black points, blue points, and orange points represent 0VP, SLVP, and LLVP, respectively. }
    \label{fig:Eamp_dep}
\end{figure*}

We further investigate dependence of VPs generation-recombination processes on $\omega_{EM}$ and the amplitude of $\Phi_B$. In~\cref{fig:Eamp_dep}, we present a phase diagram in terms of generation of long-lived VP cycles (LLVP), i.e., periodic occurrence of VPs, short-lived VP cycles (SLVP), i.e., occurrence of only a few of VP cycles, and zero VP cyles (0VP), i.e., no occurrence of VPs. In the range we demonstrate, there is no generation of VP cycles at any frequency if the amplitude of $\Phi_B < 2 \Phi_0$, and generation of periodically occurring LLVP cycles is impossible for $\tau_{EM}<2 \tau_{GL}$. While we cut our numerical investigation at $\Phi_B=10\Phi_0$, we expect that for sufficiently strong intensity of the beam, i.e., sufficiently large amplitudes of $\Phi_B$, the SC state inside the region potentially available for the generation of VPs will be completely destroyed. Additionally, for larger amplitudes of $\Phi_B$, VPs might be generated on the sides of $x=0$ line, like the snapshot of inset at $4 \tau_{GL}$ in~\cref{fig:osc_fast}(b).

While the results might visually resemble the formation and movement of kinematic vortices induced by large currents in thin films of intermediate width~\cite{berdiyorov2014dynamics, sivakov2003josephson, andronov1993kinematic}, our data also support the conjecture about Abrikosov's mechanism of vortex formation.

\subsection{\label{sec:fund1_2}Imprint profile and edge effect} 

In the previous section we considered cases where the spot size $2 w_0$ was significantly smaller than the sample size and larger than but comparable to $\xi$. In this section we change $w_0$ and report some basic observations.



\begin{figure*}
    \centering
    \includegraphics[width=0.9\textwidth]{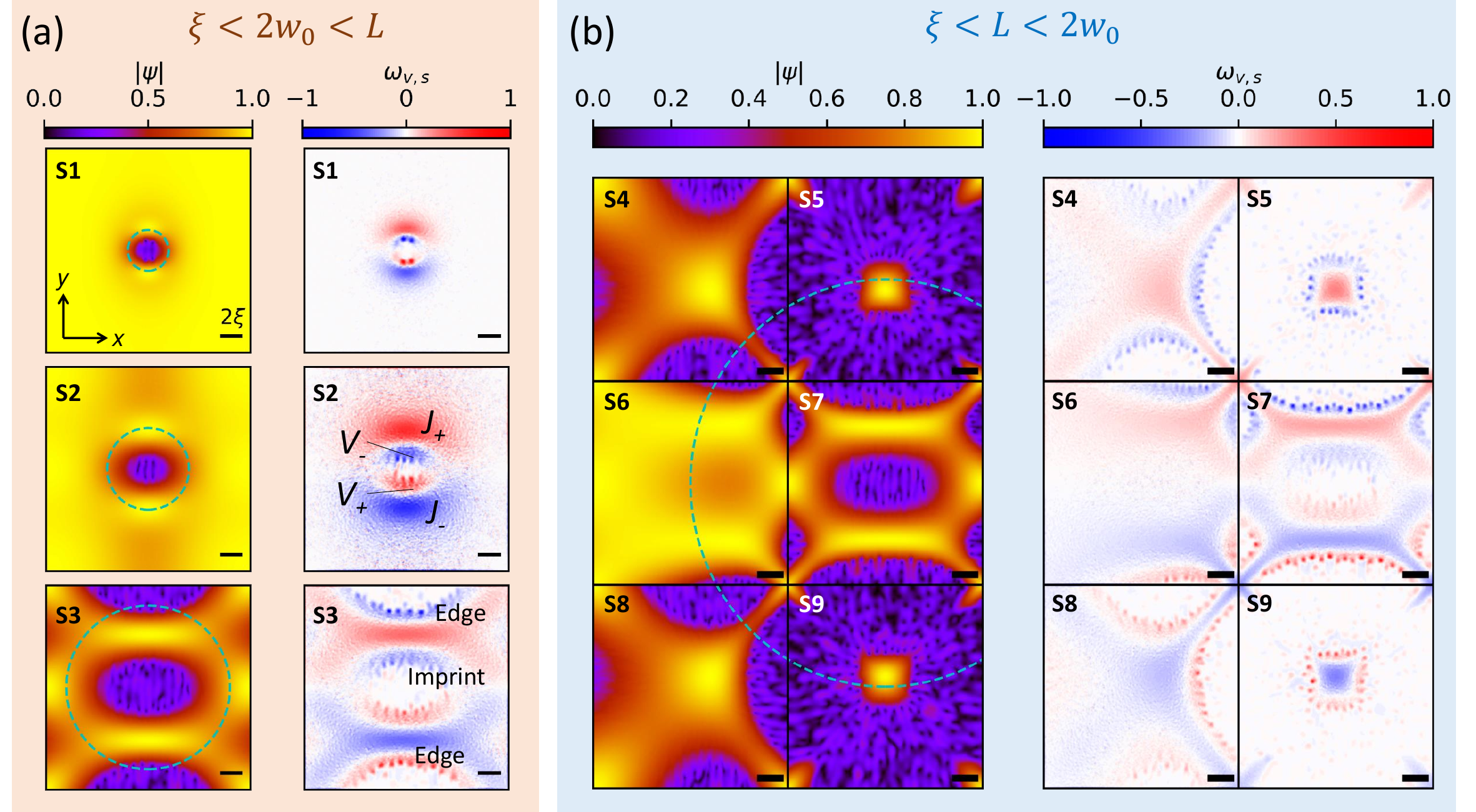}
    \caption{The distribution of supercurrent and vortex depends on different light spot sizes $w_0$, marked by green dashed circles in each sample. The $|\psi|$ shows the profile of SC states, and $\omega_{\nu,s}$ exhibits supercurrent and vortices with faint colored area and colored dots, respectively.
    (a) demonstrates cases where $w_0$ is smaller than size of sample $L$(=$20\xi$). The $w_0$ of samples S1-S3 are respectively $2\xi$, $4\xi$, and $8\xi$. The inset axis in S1 indicates direction of coordinate for all samples.
    The notations $J_+$, $J_-$, $V_+$, $V_+$ marks the supercurrent and vortices, as mentioned in the main text. The vortex clusters in S3 labeled ``Edge'' are vortices generated at the edge, and the one labeled ``Imprint'' indicates the optical imprint led generation.
    (b) Presents the cases where $2w_0>L$. The $w_0$ for all samples is $20 \xi$. Six SC samples, S4-S9, are positioned at different location under the same light spot. Those six samples are simulated independently. All results in this figure are snapshots at 1 $\tau_{GL}$ with light source $\omega_{EM} = \omega_{GL}/8$ and fixed $E_{amp}$.}
    \label{fig:size}
\end{figure*}

In~\cref{fig:size} (a), we consider case when $\xi<2 w_0<L$ for three values of $w_0$: $w_0=2\xi$ with peak magnitude of $\Phi_B= 6.4 \Phi_0$ (S1), $w_0=4\xi$ with peak magnitude of $\Phi_B= 12.8 \Phi_0$ (S2), $w_0=8\xi$ with peak magnitude of $\Phi_B= 50.9 \Phi_0$ (S3). All snapshots for $\abs{\psi}$ and $\omega_{\nu,s}$ in~\cref{fig:size} are at $t=2\tau_{GL}$, and the frequency for all cases is $\omega_{EM}=\omega_{GL}/8$.

In case S1, two VPs are generated on the sides of $x=0$. In case S2, the flux increases twice, and there are three VPs generated at $t=2\tau_{GL}$. In case S3, the distance from the edge of the spot to the edges of the film is $2 \xi$. $\Phi_B$ is eight times larger than in case S1, and the maximum and minimum of $B_z$ are located at $y=\pm 8\xi/\sqrt{2} \approx 5.66 \xi$. This results in a large magnetic flux in the areas close to the edges at $y=\pm 10 \xi$. Consequently, in combination with reduction of the barrier for a vortex to enter from an edge, vortices and antivortices are generated close to these edges as well as a number of VPs in the center of the beam.
  
Accordingly, if the beam spot is close to the edge, we observe the vortex generation at the edges even if the region is outside of the beam spot. We interpret this as a combined effect of the relatively large magnetic field close to the edge and reduction of the barrier for a vortex to enter from the edges into the sample. We also note that overall neutrality of the structure is preserved, i.e. the total vorticity of the sample is zero.

In the figure of vorticity, the profile of vortex ($V_+$), antivortex ($V_-$), supercurrent with positive/negative curl ($J_+$/$J_-$) could be clearly discerned, as indicated in Fig.~\ref{fig:size}(a). If $2w_0<L$, the distribution of vortex at $2\tau_{GL}$ predominantly shows $V_-$ and $J_+$ at upper semicircle, and vice versa. 
Similarly, with different coherence lengths, even if the ratio of $\xi$ to $\lambda_{L}$ varies, if $\xi$ is smaller than $2 w_0$, then the distribution of vortices follows the results when changing the spot size (see ~\cref{app:xi}).

In Fig.~\ref{fig:size}(a), the samples S1 and S2 exhibits spot size $2 w_0$ sufficiently small from the edge, resulting in distribution patterns that are nearly identical, except for variations in vortex numbers. In the case of S3, the distance between the spot and edge of sample is merely 2$\xi$. The order of $J_+$, $V_-$, $V_+$, $J_-$ from top to down is also broken. There are two more noticeable vortex clusters appearing at the upper and lower side, and  at the upper semicircle, which is induced by the edge of SC. The cluster size is roughly $6\xi$.

In~\cref{fig:size} (b), we consider the case when $2w_0 > L$. We shine a Gaussian beam with $w_0=20\xi$ onto six thin film samples S4-S9 positioned with respect to the beam as depicted in the figure. We note that the samples are not connected and treated independently in the simulations.  The frequency  $\omega_{EM}=\omega_{GL}/8$ is the same as for subfigure (a), and the amplitude of $\Phi_B= 480 \Phi_0$. 

The overall vorticity still roughly follows the order of $V_+$, $J_-$, $J_+$, $V_-$ from top of the beam to bottom, and accompanies with several vortex clusters at the edges. 
In the sample S7, we observe VPs with additional vortices and antivortices entered from the edges, while in the rest of the samples, we see only vortices/antivortices entered from the edges.

\section{Discussion}
\label{sec:discussion}

\subsection{Applicability of gTDGL mode and continuous light source in free space}
\label{sec:applicability}

The gTDGL equations are justified microscopically only within very small range of temperatures around $T_c$. Nevertheless, it often gives qualitatively correct results at much lower temperatures. Very close to $T_c$, microscopic calculations put certain bounds for validity of the gTDGL theory. In the time domain, changes in superconducting condensate should happen on the time scale much slower than $\tau_E$, inelastic electron-phonon scattering time~\cite{watts1981nonequilibrium,kopnin2001theory}. I.e., the theory is microscopically valid as long as $\tau_{GL} \gg \tau_E$. 
In clean samples, $\tau_E^{-1} \sim k_B T^3 / \hbar \Omega_D^2$, where $\Omega_D$ is the Debye temperature. For low-temperature superconductors such as Nb, it is typically a few GHz. Superconductors like Nb$_3$Al and MgB$_2$ can have $\tau_E^{-1} \sim 10$ GHz, and for cuprates $\tau_E^{-1}$ might be $\sim 100$  GHz. However, for dirty samples $\tau_E^{-1}$ might be larger~\cite{sergeev2000}. Experiments on dirty NbN thin films report $\tau_E^{-1} \sim 100$ GHz~\cite{sidorova2020}. 

Application of TDGL theory to driven processes can be justified for frequencies satisfying $\omega_{EM} \tau_{GL}(T) < 1/\epsilon$~\cite{plastovets2022all}, which implies
\begin{align}
    \omega_{EM} \tau_{GL,0}<1.
\end{align}

Let's represent the spot size in terms of $\lambda_{EM}$ as $w_0 = \alpha_{\lambda}\lambda_{EM}$, where $\alpha_{\lambda}$ is a real number greater than $1/2$ due to the diffraction limit in free space, and in terms of $\xi(T)$ as $w_0 = \alpha_{\xi} \xi$, where $\alpha_{\xi}>0$. In the simulations presented above, $\alpha_{\xi}$ ranged from $3$ to $20$. We then can write for the zero temperature Ginzburg-Landau coherence length ($\xi_0 \approx \xi \sqrt{\epsilon}$)
\begin{align}
\label{eq: xi0}
    \xi_0 \approx \frac{\alpha_\lambda}{\alpha_\xi} \sqrt{\epsilon} \frac{2 \pi c}{\omega}.
\end{align}

Using~\cref{tab:materials}, for example for Nb, determining maximum $\epsilon_{max}$ below which gTDGL is still microscopically valid from $\tau_{GL}/\epsilon_{max} = \tau_E$. We find $\epsilon_{max} \approx 10^{-4}$ (see~\cref{tab:materials} for other SC materials). Then for $\omega_{EM} = 10$ THz, we find at $\epsilon=10^{-4},10^{-6}$ $\alpha_\lambda / \alpha_\xi=0.02,0.2$, respectively. And, correspondingly, for $\alpha_\lambda = 1/2$, $\alpha_\xi \approx 24.8, 2.48$. Analogically for Pb, for $\omega_{EM}=4$ THZ and $\epsilon = 10^{-4},10^{-6}$, we find $\alpha_\xi \approx 28, 2.8$, respectively. However, using as the criterion for inducing LLVP cycles $\omega_{EM} \tau_{GL}(T) < 2 \pi/5$ (see~\cref{fig:Eamp_dep}, LLVP happens for $\tau_{EM}$ greater than $ 5\tau_{GL}$), we conclude that at these frequencies only SLVP cycles are possible. Using~\cref{eq: xi0}, the criterion $\omega_{EM} \tau_{GL} < 2 \pi/5$ can be rewritten as
\begin{align}
\label{eq: LLVP critertion}
    \frac{\xi_0}{\tau_{GL}} > v_0=\frac{\alpha_\lambda}{\alpha_\xi} \frac{5c}{\sqrt{\epsilon}} = \frac{\alpha_\lambda}{\alpha_\xi} \frac{1500}{\sqrt{\epsilon}} \frac{\mu\mathrm{m}}{\mathrm{ps}}.
\end{align}
For Pb, at $\epsilon = 4\cdot10^{-3}$ and $\alpha_\lambda = 1/2$, to satisfy this criterion, $\alpha_\xi$ must be larger than $10^4$, which is far from the values of $\alpha_\xi$ for which we presented simulations. 

While it is hard to satisfy criterion ~\cref{eq: LLVP critertion} in the range of microscopic validity of gTDGL, we emphasize that factor 1500 is derived from the simulations for $\alpha_\xi = 3$, $\gamma=10$, and $u=5.79$. The numerical factor can be different for different set of $\alpha_\xi$, $\gamma$, and $u$. Our observations is that for $\gamma=0$, like the standard TDGL, while the winding of order parameter in the phase is present, the amplitude of the order parameter is significantly suppressed for the range of frequencies presented in~\cref{sec:osci}, making it impossible to claim that there are any VPs, as result of ~\cref{fig:relax}(c) in~\cref{app:relaxation_time}. In this regard, we note that $\gamma \sim \sqrt{\epsilon}$, and it is possible that larger values of $\gamma$ weaken the criterion~\cref{eq: LLVP critertion}.

From the above discussion, it follows that in the region of microscopic validity of gTDGL model the LLVP phase in~\cref{fig:Eamp_dep} is harldy achievable for the normal optical techniques such as the conventional optical microscopy technique in free space.
However, we see that as temperature lowers down (i.e., $\epsilon$ increases), the criterion for reaching LLVP phase weakens. Investigation within a more exact model for lower temperatures is thus needed to answer the question about possibility reaching LLVP phase without indeed of luxury experimental equipment mentioned in next section.

\begin{table*}[]
\centering
\caption{Properties of conventional SC materials. The symbols follow the list in Table~\ref{tab:symbol}. The parameters $\xi_0$, $\lambda_{L,0}$ (or $\Lambda_0$ for thin film), $r_n$, $v_F$, $\Delta_0$. $\epsilon_{max}$, $\gamma_{max}$ are referenced from several references~\cite{kittel2018introduction, lv2021realization, black1996spectroscopy, watts1981nonequilibrium, ashcroft2022solid, townsend1962investigation, xing2008spontaneous, verma2021superconducting, bonnet1967new, gauzzi2000scaling, lee1993penetration, zverev1998anisotropy, hass1992sharp, lemberger2007penetration, draskovic2013measuring}, and $\tau_{GL,0}$, $\omega_{GL,0}/2\pi$, $\xi_0/\tau_{GL,0}$ are calculated in current work. The parameter $r_n$ is referenced from the resistivity near $T_c$.  Note that the time scale $\tau_{GL,0}$ is related to relaxation time of supercurrent~\cite{kopnin2001theory}, which is much shorter than the relaxation time of order parameter, especially for gTDGL model (see ~\cref{app:relaxation_time}).}
\begin{tabular}{llccccccccccc}

\hline
SC type 
& Material 
& \begin{tabular}[c]{@{}l@{}}$\xi_0$ \\ (nm)\end{tabular} 
& \begin{tabular}[c]{@{}l@{}}$\lambda_{L,0}$ or $\Lambda_0$\\ (nm)\end{tabular} 
& \begin{tabular}[c]{@{}l@{}}$r_n=1/\sigma$\\ ($\mu\Omega\cdot$cm)\end{tabular} 
& \begin{tabular}[c]{@{}l@{}}$\tau_{GL,0}$\\ (fs)\end{tabular} 
& \begin{tabular}[c]{@{}l@{}}$\omega_{GL,0}/2\pi$\\ (THz)\end{tabular} 
& \begin{tabular}[c]{@{}l@{}}$\tau_{E}^{-1}$\\ (GHz)\end{tabular} 
& \begin{tabular}[c]{@{}l@{}}$\xi_0/\tau_{GL,0}$\\ ($\mu$m/ps)\end{tabular} 
& \begin{tabular}[c]{@{}l@{}}$v_F$\\ ($\mu$m/ps)\end{tabular} 
& \begin{tabular}[c]{@{}l@{}}$\Delta_0$\\ (meV/THz)\end{tabular} 
& $\epsilon_{max}$
& $\gamma_{max}$ \\
\hline

\multirow{2}{*}{Type I}  

& Al & 1600  & 16 & $\sim$0.5$^{*}$ & 15 & 64 & 0.02 & 25 & 2.0 & 0.175/0.04 & $4\cdot10^{-9}$ & 340 \\ 
& Pb & 83 & 37 & $\sim$2.3$^{**}$ & 75 & 13 & 50 & 1.1 & 1.8 & 2.8/0.7 & $4\cdot10^{-3}$ & 17 \\ 
\hline

\multirow{3}{*}{Type II} 

& Nb   & 38  & 39  & 2$^{\dag}$  & 96  & 10 & 1,4 & 0.4 & 1.4 & $\sim$2.5/0.60 &  $1\cdot10^{-4}$ & - \\ 
& YBCO & 1.4 & 210 & $\sim$150$^{\ddag}$ & 37 & 27 & 690 & 0.04 & $>$0.7 & 20/4.8 & 0.025 & - \\ 
& Nb (2 nm)   & 10-100  & $10^3$-$5\cdot 10^5$  & 153  & 1.6-820  & 1-600  & - & 0.1-60 & - & 0.18 & - & - \\ 
\hline

\multicolumn{12}{l} 
{\footnotesize  $^{*}r_n$ of 300-nm Al. \newline $^{**}r_n$ of 100-nm Pb (dimensional $r_n$ shown in the preprint version~\cite{xing2008spontaneous_pre}). \newline $^{\dag}r_n$ of 160-nm Nb. \newline $^{\ddag}r_n$ in the $ab$-plane. }

\end{tabular}

\label{tab:materials}
\end{table*}

Another aspect that needs discussion is heating of the SC sample due to irradiation of light source. The power of Gaussian beam is 
\begin{align}
\label{eq: Power output}
    P=\frac{\pi (E_0 w_0)^2}{4 \mu_0 c}.
\end{align}
In~\cref{sec: Results}, we have seen that flux about $2 \Phi_0$ is needed to produce generation of a VP. Using~\cref{eq:Phi_B,eq: Power output}, the power of the beam that produces $\Phi_B=\Phi_0$ is
\begin{align}
    P=\frac{\pi (\Phi_0 \omega_{EM})^2}{8 \mu_0 c}.
\end{align}
For $\omega_{EM}=1,10$ THz, $P=4.45, 445$ nW, respectively. For the spot size $w_0 \sim \lambda$, the corresponding intensity of radiation is $\sim 0.445, 44.5$ nW/$\mu$m$^2$. As suggested in~\cite{plastovets2022all}, one can think about using a substrate to dissipate this power from the area of the beam spot. For a sapphire substrate of thickness about $1$ $\mu$m and thermal conductivity $\kappa ~\sim 1000$ W/(m$\cdot$K)~\cite{berman1955} and temperature difference between the substrate and the SC sample $\Delta T = 0.001$ K, the density of the heat transfer is $\sim 1$ $\mu$W/$\mu$m$^2$.

\subsection{\label{sec:optic}Proposed experimental setup}

As mentioned in~\cref{sec:applicability}, considering on the experiment with continuous wave of light source, the narrow range of $\epsilon$ push the experiment in the harsh situation, i.e., $T_c-T<10^{-4} T_c$. Nonetheless, even if the applicable range of gTDGL model is extremely narrow, it is still used to simulate the vortex dynamics and provide the phenomenological information. 
For instance, the prediction of kinematic vortex within gTDGL model~\cite{andronov1993kinematic} found its confirmation in  experiments~\cite{sivakov2003josephson}. 
The cutting edge light sources, such as the techniques of few cycle-pulse laser technique~\cite{zhang2021intense, sell2008phase} and the spot size breaking through the diffraction limit\cite{wimmer2014terahertz, yoshida2019subcycle, gadalla2014design, tuniz2023subwavelength}, has been developed over decade, and beyond several order of the restriction of the criteria. The~\cref{eq: LLVP critertion} can therefore take account the situation $\omega_{EM} \tau_{GL} \geq 2 \pi / 5$ for the former technique and $\alpha_{\lambda}<$1/2 for the latter one.

Additionally, we also point that increasing energy deposited by the beam on a SC may induce thermal effects, which are undesirable in this context. Thus, one would need to come up with the mitigation strategy to negate the effects of heating.
We therefore see that to take advantage of the engineering of VP states and complex phase dynamics one would need to carefully balance these constraints between the SC and the light. This in itself is not an insurmountable obstacle but clearly the protocol would depend on specific materials and details or optical technique. 
The application of aforementioned light sources also efficiently alleviate the thermal effect. The pulse laser technique which is known for cold processing of surgeries and materials ~\cite{yan2022femtosecond, he2024metal}, and sub-wavelength spot size reduce the heating area directly. 

In this section, we propose candidates of experimental configurations that correspond to the scenario of ~\cref{fig:osc_fast}(a) based on these two techniques. 
Let us focus on a specimen of 2 nm niobium thin film at $T=0.99T_c$ with $\Lambda(T)=10 \mu$m (corresponding to $\lambda_{L,0}=45$nm), thin film resistivity $r_n = 153$ $\mu \Omega \cdot$ cm, and $\tau_{GL}(T) = 160$ fs ($\omega_{GL}/2\pi\approx $6 THz). In the context of niobium thin films, the Pearl length, which serves as the effective penetration depth, can range from 1 to 500 $\mu$m when the thickness is 2 nm~\cite{lemberger2007penetration, draskovic2013measuring}. So, we expect that this long penetration depth maintains the vortex dynamics similar to our simulation, which is without considering the screening effect. 
Since the lower the frequency $\omega_{EM}$ the more light-induced VPs are generated(see~\cref{fig:Eamp_dep}), we consider the case $\omega_{EM}=\omega_{GL}/10 \approx 0.6$ THz for thin niobium film, like the scenario of ~\cref{fig:osc_fast}(a). It corresponds to a minimum spot size of 250 $\mu$m, which is at the diffraction limit of experimental conditions in free-space. 

The induced $B_z$ of linearly-polarized light depends on the non-uniform distribution of light intensity. Thus, $B_z$ increases significantly when the spot size is reduced to less than $\frac{1}{2}\lambda_{EM}$. It can be implemented by using the near-field THz radiation. Additionally, keeping the photon energy below the SC gap with a small spot size is advantageous as it avoids inducing quasiparticles and does not lead to serious thermal effects. 
For the case of $\omega_{EM}=\omega_{GL}/10$ and maintaining the same spot size as in the simulation (0.6 $\mu$m), the estimated $E_{amp}$ is $30.4$ kV/cm. These lower frequencies will help avoid photo-excited carriers crossing $\Delta_0$. Nowadays, state-of-the-art light sources can achieve transient terawatt energy levels~\cite{backus1998high} and 100 MV/cm of THz electric fields~\cite{zhang2021intense, sell2008phase}. If we consider the intense THz technique  combining the sub-wavelength spot size, the technique of near field nano-tip THz photoemission which is usually applied by the scanning near-field optical microscopy (SNOM), the transient and local THz electric field can achieve orders of amplitude from kV/cm to MV/cm~\cite{wimmer2014terahertz, yoshida2019subcycle}. For the observation of vortices, the size of the tip is determined by the scale SC coherent length, which is expected to maintain the sub-micrometer spot size. Instead of utilizing the tip as a probe in SNOM, we consider it as a THz pump for VP generation. 
Another approach to surpass the diffraction limit involves the techniques of metamaterials~\cite{gadalla2014design}, near field THz antenna~\cite{tuniz2023subwavelength}, etc. Furthermore, if spot size in this regime is comparable to the coherence length, presenting substantial potential for applications such as single vortex trapping.

In a few-cycle THz pulse techniques, the observation of LLVP phase is impossible as the number of light cycles is limited. But the scenarios similar to the simulations illustrated in~\cref{fig:osc_fast}(b) and ~\cref{fig:relax} are observable. In such a case, the minimum $2w_0$ is considered to be 0.5 $\mu m$ for $\omega_{EM}=\omega_{GL}/2$ in simulation. To maintain a constant flux condition, specifically a peak value of $\Phi_B\sim 9.6 \Phi_{0}$ (similar as the case of Fig.~\ref{fig:osc_fast}(b)), the required $E_{amp}$ is approximately 448 kV/cm, which is achievable with intense THz techniques~\cite{zhang2021intense, sell2008phase}. In ~\cref{app:relaxation_time}, the simulation of sub-cycle optical pulse also demonstrate the generation of single-cycle VP.

On the probe side, $\xi$ and $\tau_{GL}$ are associated with the range of time and space resolution of instrument, which can be observe by time-resolved scanning SQUID~\cite{cui2017scanning} and time-resolved scanning tunneling microscopy~\cite{sheng2022launching}.

\section{\label{sec:summary}Summary}

We propose the new approach to quantum steering of a coherent electron state by Gaussianly structured light. The non-uniformity of the vector potential of linearly-polarized Gaussian beam leads to a non-zero z-component of the magnetic field antisymmetic with respect to the coordinate in the direction perpendicular to the polarization of the electric field. The maximum and minimum of the magnetic field are located away from the center of the beam and are the spots where the amplitude of time dependent magnetic filed is maximal. While total magnetic flux is zero, upper and lower half-planes in the plane of the film are threaded by oscillating fluxes of different sign. This creates a situation where the barrier to introduce a vortex into the bulk can be overcome by generation of a vortex-antivortex (VP) pair. During half-period of the EM wave, vortices in such VPs move towards the position of maximum/minimum of the magnetic field and back to recombine at the beam's center. In this approach, the light induces dynamical vortex state with zero total vorticity and can transfer quantum numbers (like spin and orbital angular momentum) to the superconducting condensate, which will be illustrated in the follow-up paper. This process can be viewed as the quantum printing where we imprint the quantum numbers of the light field.



Based on gTDGL theory, we show that depending on the frequency of the incident linearly-polarized Gaussian beam, the generation of VPs may occur in a periodic fashion (LLVP phase) or can last  for a few cycles (SLVP phase). Although in the region of microscopic validity of gTDGL for the ratios of the spot size and coherence length we were able to demonstrate in simulations LLVP phase is not achievable, the criterion for LLVP phase weakens as temperature lowers down, suggesting that LLVP phase for rather small ratio $w_0/\xi$ might be achievable at lower temperature within more exact theory. Simulations for larger ratios $w_0/\xi$ within the range of validity of gTDGL model are required to conclude about possibility of LLVP phase at temperatures close to $T_c$.

Although the vortices in VPs induced by the beam generate and recombine, one can perhaps prohibit their recombination via separating them using the Hall effect in the presence of a current flowing through the film perpendicular to the motion of vortices in VPs or employing the inverse Faraday effect in superconducting condensate~\cite{mironov2021,plastovets2022all}, potentially leading to controlled generation of vortices with EM radiation.

We also discuss a broader context of presented work. Generation of vortices under light is expected on general grounds. The light 
provides transient electric and perpendicular magnetic fields which in turn generate  and  move vortices in vortex-antivortex pairs. The approach is similar to the printing of matter where the recipient substance receives the ink to produce the copy. 

In our approach, the light can transfer energy and create oscillating magnetic fields. Light also  can transfer quantum numbers (like spin and orbital angular momentum of the beam) to SC  and induce new patterns of vorticity.  These results  will be presented in the follow up paper. Overall, this process of unduction of topological excitations in quantum fluid  can be viewed as the {\em quantum print} where we imprint the quantum numbers of the light field on coherent electron fluid - the SC .

\section{\label{sec:acknowledg}Acknowledgements}

Part of this work was carried out while AB and his collaborators were visiting Stanford Department of Physics and SLAC. We are grateful to T. Deveraux and members of his group for discussions and hospitality during our visit.  We also acknowledge useful discussoins with  K. Bondarenko, D.H. Lee, R.B. Laughlin,  G. Jotzu, A. Pathapati, S. Bonetti,  J. Wiesenrieder, M. Manson, O. Tjernberg, P. Wong, Z.X Shen, H. Hwang, B. Friedman, Y. Suzuki, W. Lee, A. Metha, A. Chainani and  M. Verma. 

We acknowledge support from the European Research Council under the European Union Seventh Framework ERC-2018-SYG 810451 HERO, Knut and Alice Wallenberg Foundation KAW 2019.0068, and the University of Connecticut.

\appendix

\section{\label{app:Jc}Critical current density $J_c$} 
In gTDGL approach the  driven current  can be used to unpair Cooper pairs when it is higher than typical critical current, which is usually a few $0.1 J_0$~\cite{watts1981nonequilibrium, berdiyorov2014dynamics}. We present here the simulation that clearly demonstrates the presence of the critical current density of the SC thin film in our simulation.
\begin{figure}[h]
    \centering
    \includegraphics[width=0.35\textwidth]{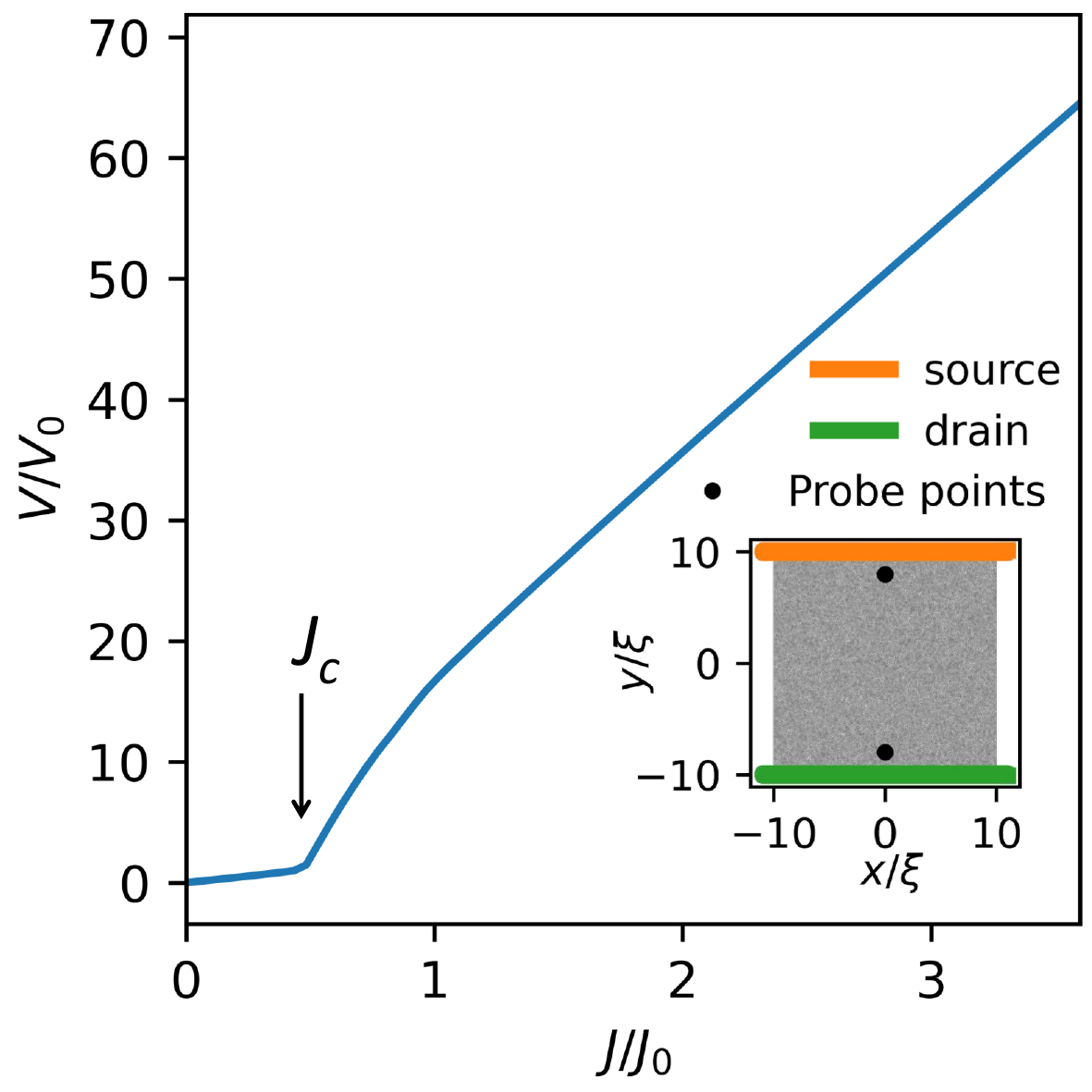}
    \caption{The I-V curve of SC thin film. The $J_c$ is estimated as $0.48 J_0$.
    }
    \label{fig:Jc}
\end{figure}

 \section{\label{app:relaxation_time}Relaxation time of healing process of $\abs{\psi}$}

 \begin{figure*}
    \centering
    \includegraphics[width=0.85\textwidth]{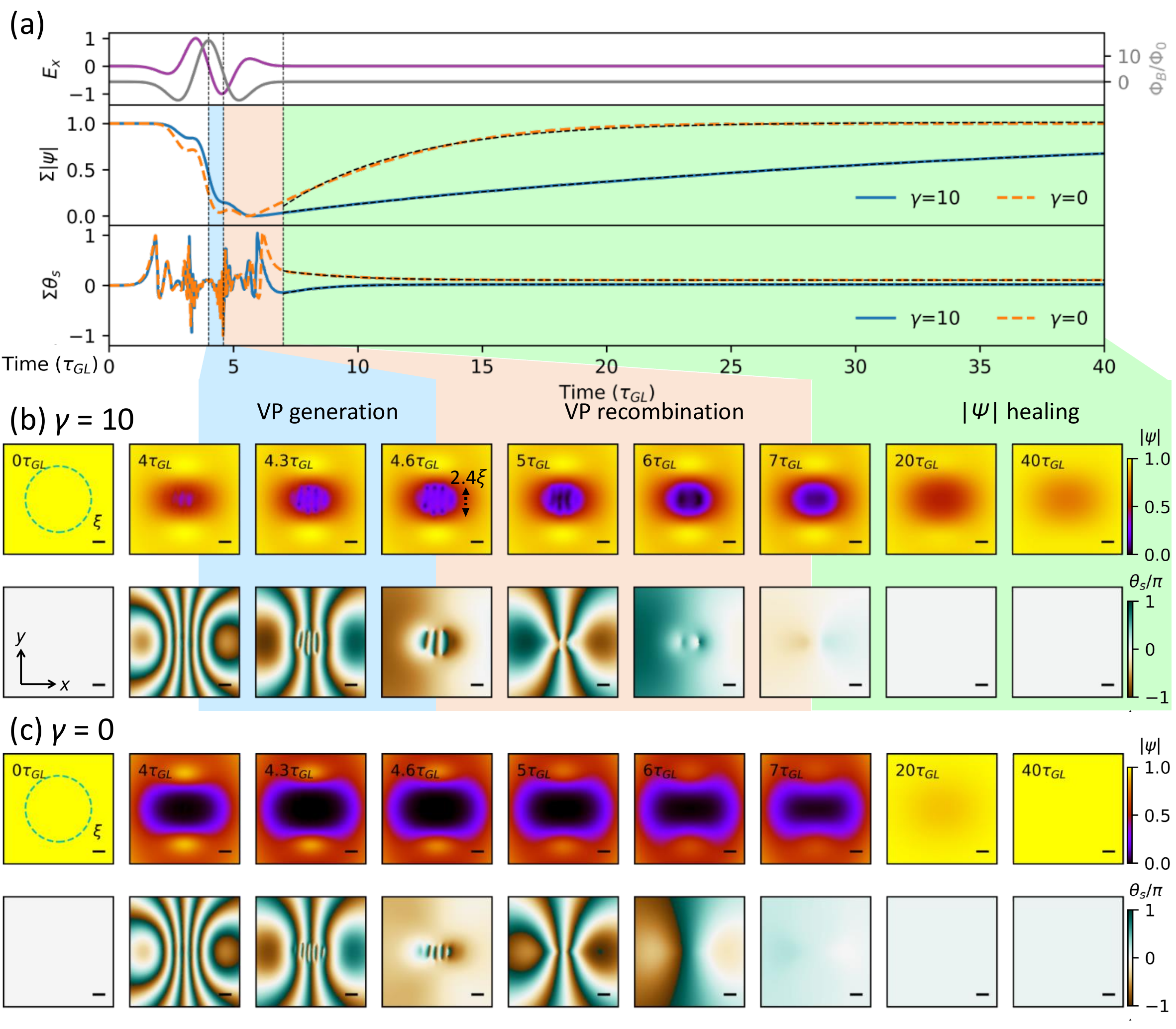}
    \caption{The time evolution of $|\psi|$ and $\theta_s$ with optical pulse. 
    (a) Top subplot shows normalized $E_x$ and $\Phi_B$ of the optical pulse. The optical pulse $E_x$ is generated by the third derivative of the Gaussian distribution, and $\Phi_B$ proportional to integral of $E_x$, is the second time derivative of the Gaussian distribution.
    The following two subplots represent total $|\psi|$ (denoted by $\sum|\psi|$), and summation $\theta_s$ ($\sum\theta_s$) across the entire area of sample, respectively. The $\sum|\psi|$ curve is obtained by the min-max normalization, and the $\sum\theta_s$ curve is calculated via the max absolute scaling. The blue solid lines and orange dashed lines are represent the SCs with $\gamma=10$ and $\gamma=0$, respectively. 
    The time-evolution can be categorized into 3 processes: VPs generation  from 4.0-4.6 $\tau_{GL}$ (blue region), VP recombination from 4.6 to roughly 7 $\tau_{GL}$ until the last VP recombination (pink region), and the healing process of the order parameter after 7 $\tau_{GL}$ (green region).
    The relaxation time of the healing process is determined by an exponential fit $c_1 \exp{(-t/\tau_{healing})}+c_2$ after 7 $\tau_{GL}$ with fitting constants $c_1$ and $c_2$, and the fitting curves are shown as the black dashed lines. 
    (b) and (c): The selected zoom-in snapshots of $|\psi|$ (upper row) and $\theta_s$ (lower row) for demonstrating different stages of (b) $\gamma=10$ and (c) $\gamma=0$. The green dashed circles and inset axis at the left figures illustrate the beam spot and direction of coordinate. }
    \label{fig:relax}
\end{figure*}

In this appendix, we proceed to investigate the relaxation time of healing process of $\abs{\psi}$ by the optical pulse. From the results of typical light-VP interaction in~\cref{sec:osci}, the intrinsic time duration of generation, recombination, and healing delay the synchronizing with light oscillation. For instance, the VP cycle behaves almost ``in-phase'' with amplitude of $\Phi_B$ when $\omega_{EM}=\omega_{GL}/40$, but it turns to be ``out-of-phase'' when $\omega_{EM}=\omega_{GL}/10$. The approach of single-cycle optical pulse induced VP generation enables us to study the equilibrium process with minimal intervention and provides insight into the appropriate off-peak interval of the vector potential. 

Here, we utilize the second time derivative of the Gaussian distribution as our $\Phi_{B}$. The relaxation processes are present in~\cref{fig:relax}. We consider the relaxation process of two cases: $\gamma=10$ and $\gamma=0$, which represent the gTDGL model and standard TDGL model, respectively. The fitting results of relaxation time of healing, $\tau_{healing}$, are listed as follows: For the case of $\gamma=10$, $\tau_{healing}=$ 30.9 and 1.6 $\tau_{GL}$ for $|\psi|$ and $\theta_s$, respectively; and for the case of $\gamma=0$, $\tau_{healing}=$ 5.1 and 2.5 $\tau_{GL}$ for $|\psi|$ and $\theta_s$, respectively. Consequently, it comes up with the long healing process of $|\psi|$ and short for $\theta_s$. The relaxation process follows the exponential decay. The decay time of order parameter is approximately $u\cdot\tau_{GL}$ for TDGL model ($\gamma=0$) and longer for gTDGL model ($\gamma\neq 0$). The relaxation of $\theta_s$ shows the stronger coherent behavior with light.

 \section{\label{app:xi}Distribution of vortex dependent on $\xi$} 

 \begin{figure}[]
    \centering
    \includegraphics[width=0.4\textwidth]{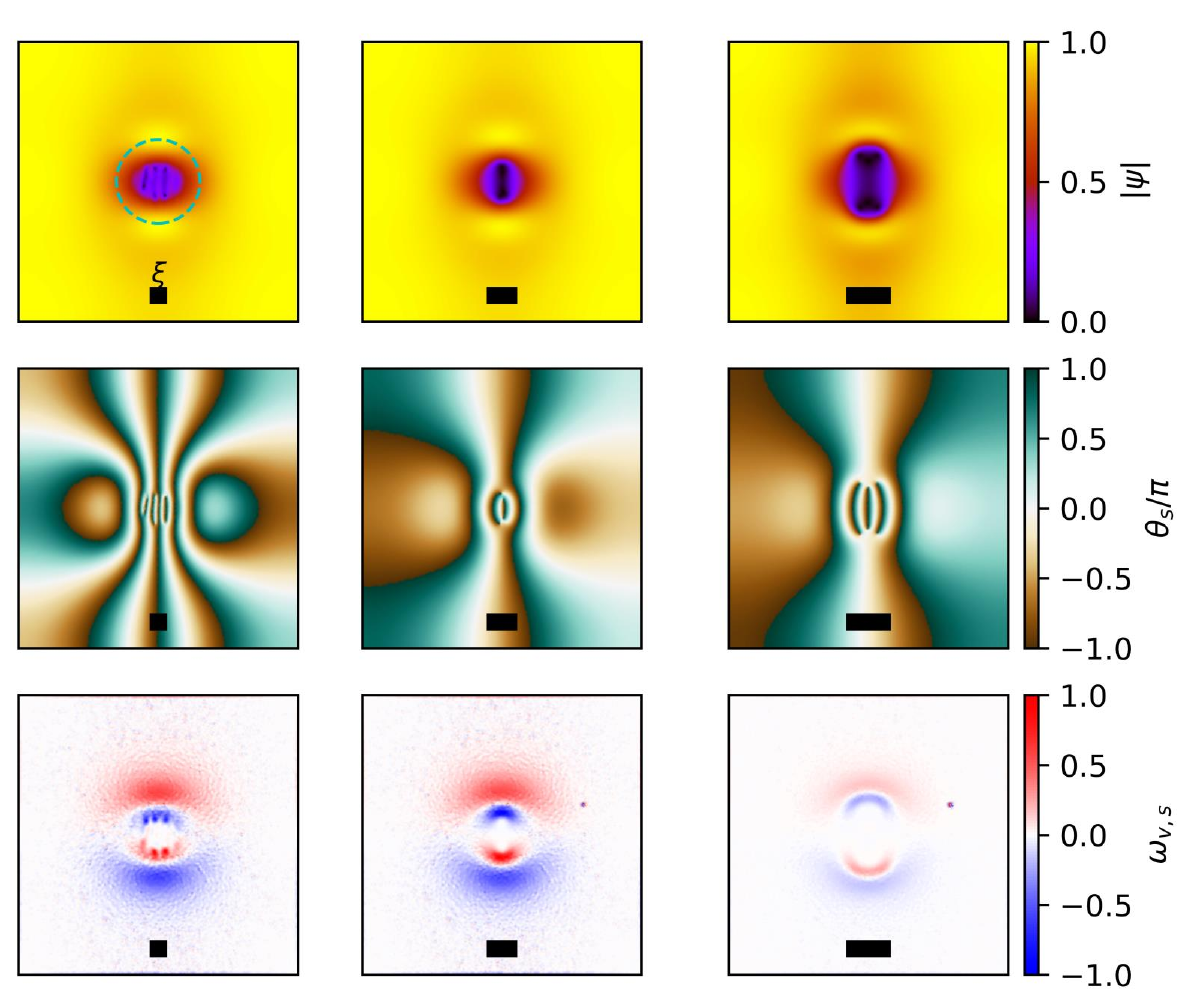}
    \caption{The simulation of same spot size with different coherence length $\xi$. The profile of $\psi$, $\theta_s$, and $\omega_{\nu,s}$ are the snapshots at 2 $\tau_{GL}$ for the light source $\omega_{EM}=\omega_{GL}/8$ and $2 w_0=0.6 \mu m$. The $\xi$ are chosen to be 0.1, 0.2, and 0.3 $\mu m$ for the first to third column, respectively.
    }
    \label{fig:xi}
\end{figure}

In this appendix, we demonstrate how the distribution of  vorticity is dependent on the coherence length $\xi$. ~\cref{fig:xi} demonstrates the profile of $\abs{\psi}$, $\theta_s$, and $\omega_{\nu,s}$ for three cases $\xi=0.1, 0.2$, and $0.3 \mu$m below the length of spot $2 w_0 = 0.6 \mu$m. In this simulation, we can observe that the distribution of vorticity are similar. This test tracks the changes on coherence length and imprinted vorticity  with same $\lambda_L$ ($0.1 \mu$m).


\bibliography{LightSC}

\begin{thebibliography}{58}
\bibitem [{sup({\natexlab{a}})}]{supp3}%
  \BibitemOpen
  \href@noop {} {} ({\natexlab{a}}),\ \bibinfo {note} {see {S}upplemental {M}aterial at URL-will-be-inserted-by-publisher for the video of Fig. 4(a).}\BibitemShut {Stop}%
\bibitem [{sup({\natexlab{b}})}]{supp4}%
  \BibitemOpen
  \href@noop {} {} ({\natexlab{b}}),\ \bibinfo {note} {see {S}upplemental {M}aterial at URL-will-be-inserted-by-publisher for the video of Fig. 5(a).}\BibitemShut {Stop}%
\bibitem [{sup({\natexlab{c}})}]{supp5}%
  \BibitemOpen
  \href@noop {} {} ({\natexlab{c}}),\ \bibinfo {note} {see {S}upplemental {M}aterial at URL-will-be-inserted-by-publisher for the video of Fig. 5(b).}\BibitemShut {Stop}%
\end{thebibliography}

\end{document}